\documentclass[%
 reprint,
superscriptaddress,
amsmath,amssymb,
 aps,
 prl,
]{revtex4-2}

\usepackage{graphicx}
\usepackage{dcolumn}
\usepackage{bm}

\begin{document}

\preprint{APS/123-QED}

\title{Multi-color Wavefront Sensor using Talbot effect for High-order Harmonic Generation}

\author{Yang Du}
\thanks{These authors contributed equally to this work.}
\affiliation{Institute of Advanced Science Facilities, Shenzhen, 518107, China}%

\author{Kui Li}%
\thanks{These authors contributed equally to this work.}
\affiliation{Aerospace Information Research Institude, Chinese Academy of Sciences, Beijing, 100094, China}
\affiliation{School of Optoelectronics, University of the Chinese Academy of Sciences, Beijing, 100049, China}

\author{Jin Niu}%
\affiliation{Aerospace Information Research Institude, Chinese Academy of Sciences, Beijing, 100094, China}
\affiliation{School of Optoelectronics, University of the Chinese Academy of Sciences, Beijing, 100049, China}

\author{Angyi Lin}
\affiliation{Department of Electrical and Electronic Engineering, Southern University of Science and Technology, Shenzhen, 518055, China}%

\author{Jie Li}%
\affiliation{Aerospace Information Research Institude, Chinese Academy of Sciences, Beijing, 100094, China}
\affiliation{School of Optoelectronics, University of the Chinese Academy of Sciences, Beijing, 100049, China}

\author{Zhongwei Fan}%
\affiliation{School of Optoelectronics, University of the Chinese Academy of Sciences, Beijing, 100049, China}

\author{Guorong Wu}%
\affiliation{State Key Laboratory of Molecular Reaction Dynamics, Dalian Institute of Chemical Physics, Chinese Academy of Sciences, Dalian, 116023, China}%
\affiliation{Institute of Advanced Science Facilities, Shenzhen, 518107, China}%

\author{Xiaoshi Zhang}
\email{To whom correspondence should be addressed. zhangxiaoshi@itc.ynu.edu.cn and zhangfc@sustech.edu.cn}
\affiliation{ School of Physics and Astronomy, Yunnan University, Kunming, 650500, China}
\affiliation{Aerospace Information Research Institude, Chinese Academy of Sciences, Beijing, 100094, China}

\author{Fucai Zhang}
\email{To whom correspondence should be addressed. zhangxiaoshi@itc.ynu.edu.cn and zhangfc@sustech.edu.cn}
\affiliation{Institute of Advanced Science Facilities, Shenzhen, 518107, China}%
\affiliation{Department of Electrical and Electronic Engineering, Southern University of Science and Technology, Shenzhen, 518055, China}%

\date{\today}

\begin{abstract}
 We present a novel method for multi-color wavefront measurement of high-order harmonic generation beams using the Talbot effect, validated both theoretically and experimentally for the first time. Each harmonic maintains a unique wavefront and produces an independent set of self-images along the optical axis.We achieved the wavefronts reconstruction of three harmonics in a single measurement scan, expanding the spectrally-resolved capability of the conventional Talbot effect wavefront sensor. This breakthrough introduces a novel tool for studying the multi-color wavefront in high-order harmonic generation, unlocking the potential to investigate spatiotemporal ultrafast nonlinear dynamics in attosecond pulse formation on a shot-by-shot basis.
\begin{description}
\item[PACS number]
42.15.Dp, 42.30.Rx, 42.60.Jf, 42.65.Ky
\end{description}
\end{abstract}

\maketitle



The radiation produced by High-order Harmonic Generation (HHG) exhibits distinct spatial features at the angstrom level and temporal characteristics at the attosecond level, spanning the extreme ultraviolet to soft X-ray spectra regimes \cite{bartels2002generation}. Understanding the spatiotemporal coupling in HHG beams is essential for decoding the multidimensional features of complex incident fields \cite{akturk2010spatio}, providing insights into atomic responses and optimizing applications such as attosecond-driven ultrafast process exploration \cite{sansone2010electron,corsi2006direct}, table-top lensless microscopy \cite{sandberg2007lensless}, beam shaping and optimal focusability \cite{sanson2020highly,hoflund2021focusing,abbing2022extreme}, and seeding free-electron lasers \cite{lambert2008injection}. Despite their importance, these pursuits face challenges arising from fundamental physical mechanisms and technical complexities.

Often, characterizing HHG beams involves averaging over temporal or spatial dimensions. Temporal diagnosis via photoelectron spectroscopy integrates across all points in the beam, and spatial wavefront measurement using Hartmann wavefront sensor or point diffraction interferometry averages over wavelengths \cite{dacasa2019single,li2020high,lee2003wave}, which may be insufficient. Recently, several novel wavefront sensing methodologies with spectral resolution capabilities have been introduced. The Spectral Wavefront Optical Reconstruction by Diffraction (SWORD) technique, involves scanning slits near the focus of HHG beams and recording diffraction signals using a flat-field spectrometer \cite{frumker2012order}. Some potential drawbacks include suitability for sources with rotational symmetry, inherent scanning properties, attenuation caused by slits, and limited spatial resolution. Another approach, lateral shearing interferometry (LSI) \cite{austin2011lateral}, relies on generating two identical beams and controlling the shear between them. However, both methods can only measure 1D signals and require rotating the measuring device 90 degrees for complete 2D wavefront measurement. A recent development is a Hartmann sensor with spectral diversity \cite{freisem2018spectrally}, where traditional absorption gratings are integrated into the aperture of the Hartmann mask, providing spectral resolution. However, wavefront reconstruction depends on identifying each diffraction peak for each aperture, necessitating a large spacing between apertures, ultimately limiting spatial resolution and reconstruction accuracy. Ptychographic wavefront sensing \cite{du2023high,liu2023observation} records a series of diffraction patterns by translating across the probe on an object in overlap regions. Through iterative phase-retrieval algorithms, both complex-valued objects and probe beams can be reconstructed simultaneously. Its unique advantage lies in robust reconstructions with significantly higher spatial resolution than Hartmann sensors, allowing the simultaneous reconstruction of multiple spatial and/or temporal modes. Nevertheless, ptychography is an indirect measurement that the fidelity of the retrieved results is sometimes questionable and needs further validation. Limitations stem from inherent scanning characteristics, prolonged exposure times, and the substantial computational iterations required for phase retrieval algorithm implementation.

The Talbot effect, also known as self-imaging, is a near-field diffraction phenomenon first observed by Talbot in 1836 \cite{talbot1836facts}. When a monochromatic coherent beam traverses a one-dimensional (1D) or two-dimensional (2D) periodic structure, such as a grating, induces diffraction in free space, resulting in the emergence of a pattern replicating the original object at an integer Talbot distance related to the wavelength.  At fractional Talbot distances along the diffraction direction, the periodic pattern reappears but its periodicity is reduced by an integer factor relative to the initial input pattern \cite{berry1996integer}. Planar wavefront illumination is usually presumed, and similar self-imaging phenomena can be observed even with Gaussian-like wavefront illumination \cite{patorski1989self}, in this case, the self-image is a lateral amplification or compression of the original ones. The unique manifestations of the Talbot effect have found applications in wavefront sensing. As a most natural and direct way to measure wavefront, the Talbot effect Wavefront Sensor (TWS) offers notable advantages. Its straightforward experimental setup reduces systematic errors, enabling precise measurements in a single exposure. This approach is particularly beneficial for characterizing beams with a large Numerical Aperture (NA, up to 0.5NA), high accuracy (up to ${\lambda/100}$), and fast feedback in various radiation \cite{siegel2001wavefront,miyakawa2015extending,yamada2020x,seaberg2019wavefront,liu2018high}. However, the mentioned applications rely on quasi-monochromatic  illumination or wavelength averaging post-processing, and there is currently remain unexplored spectrally resolved wavefront diagnosis for comprising multi-color nature and intricate nonlinear phase front such as HHG beams. 

In this letter, we present the first theoretical and experimental evidence of the high-order harmonic Talbot effect. Our findings confirm that each harmonic maintains its distinct wavefront during propagation and generates self-images at independent positions along the optical axis. This presents a new and promising single measurement scan method for decoding the multi-spectral wavefronts of the HHG.

\begin{figure}[h]
	\includegraphics[width=8.6cm,keepaspectratio]{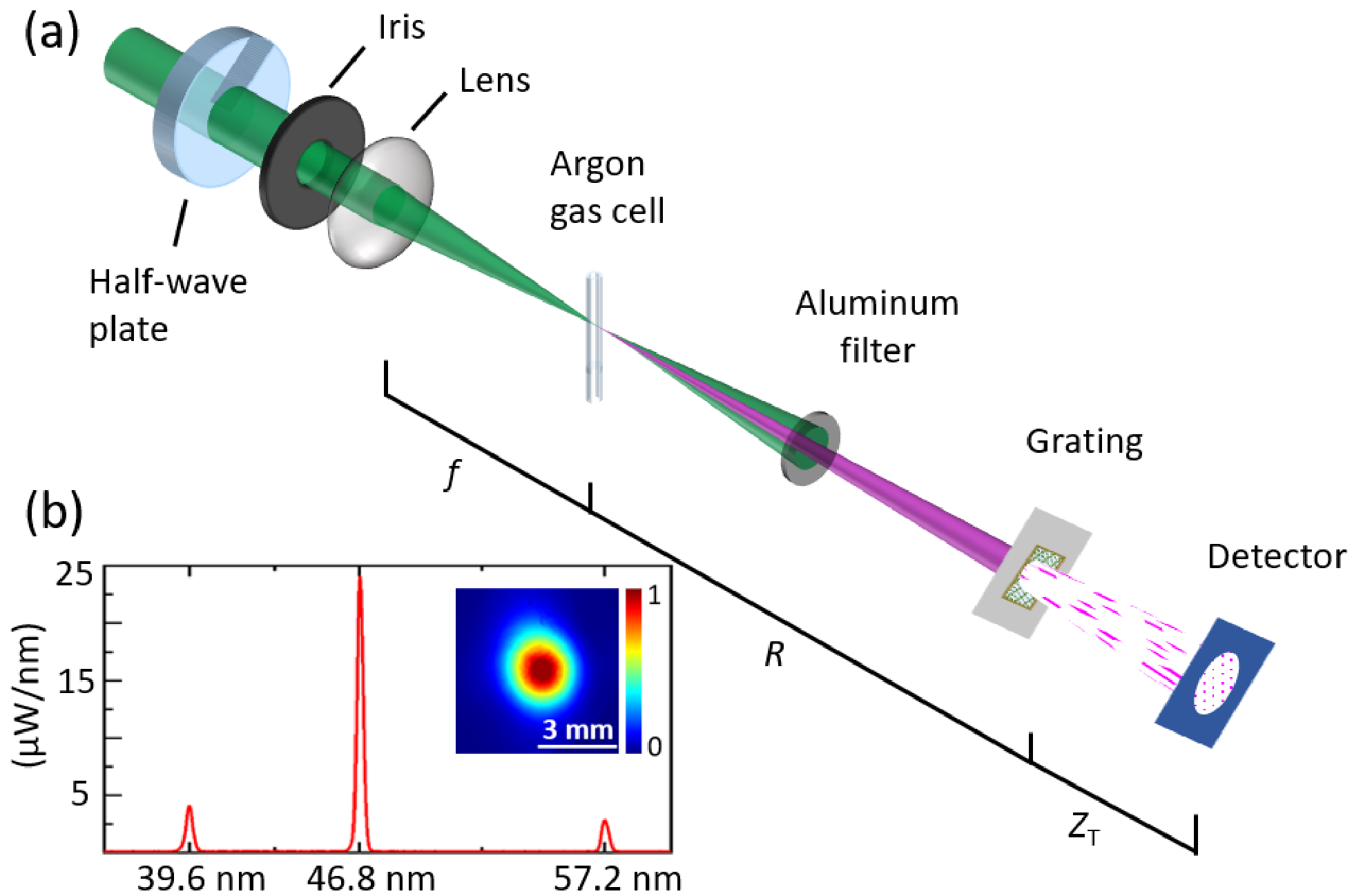}
	\caption{\label{PRL_fig_1} (color online) schematic of the Talbot effect wavefront sensor. (a) is the setup of the experiment, and (b) are the spectra and intensity characteristics of the HHG beams.
	}
\end{figure}

The TWS simply consists of a 2D grating and a detector, as illustrated in Fig.~\ref{PRL_fig_1}.

According to the Fresnel–Kirchhoff diffraction theory,  The diffracted amplitude field is given by:
\begin{eqnarray}
	u\left( {{{\vec r}_2}} \right) \propto &&\frac{{\exp \left( {i2\pi z/\lambda } \right)}}{{i\lambda z}}\exp \left( {i\phi } \right) \nonumber \\
	&&\times\int {t\left( {{{\vec r}_1}} \right)\exp \left[ {i\frac{\pi }{{\lambda z}}{{\left| {{{\vec r}_2} - {{\vec r}_1}} \right|}^2}} \right]} d{\vec r_1}\label{eq-1},
\end{eqnarray}
where $\lambda$ is the wavelength, $z$ is the propagation distance from the grating to the observation plane,  and ${\vec r}_2$ and ${\vec r}_1$ are located at the observation and grating planes, respectively. the diffracted field amplitude \emph{u}(${\vec r}_2$) is defined in terms of the transfer function of the grating and the phase of the wavefront \emph{exp}($i\phi$) at the grating plane originating from the source. For a coherent Gaussian-like illumination, the Talbot distance $Z_{\rm{T}}$, scaled by the geometrical magnification of $M$, is expressed as follows:
\begin{eqnarray}
	{Z_{\rm{T}}} = \eta \frac{{{p_0}^2}}{\lambda }M = \eta \frac{{{p_0}^2}}{\lambda }\frac{{L}}{R}\label{eq-2},
\end{eqnarray}
where $\eta$ is a positive integer or an irreducible fraction, $p_0$ is the period of the grating, $L=R+Z_{\rm{T}}$ is the distance of the focal to the imaging plane, and $R$ is the distance of the focal to the grating plane. The period of the detected self-image pattern $p_1$ is also equally magnified by:
\begin{eqnarray}
	{p_1} = {p_0}\frac{{L}}{R}\label{eq-3},
\end{eqnarray}
by synthesizing the scalar formula for the Talbot distance and incorporating geometric amplification relations, one can derive the distance from the focal point to the grating plane \cite{yashiro2009hard}:
\begin{eqnarray}
	R = \frac{L}{2}\left[ {1 \pm {{\left( {1 - \frac{{4\eta {p_0}^2}}{{\lambda L}}} \right)}^{1/2}}} \right]\label{eq-4}.
\end{eqnarray}
Equation (4) yields two solutions. One solution situates the grating in proximity to the detector, while the other places the grating close to the focus, resulting in enhanced magnification and angular sensitivity. In instances of wavefront distortion during illumination, the centroid of the achieved pattern deviates from the original grid regarding the incident beam. The shift in these vectors correlates with the wavefront gradient, retrievable through an integration process. Generally, the self-image can be regarded as a sheared copy of the original wavefront, characterized by a shear angle $\theta_s$ of $2\lambda/p$, Consequently, a smaller $p$ results in a larger shear angle, enhancing sensitivity and signal levels \cite{liu2018high}. For a single grating, consideration must be given to the pixel size of the detector, where $p$ is always consistently three times larger than the pixel size to strike a balance between sensitivity and adequate detector sampling \cite{grizolli2017single}. 

\begin{figure}[h]
	\includegraphics[width=8.6cm,keepaspectratio]{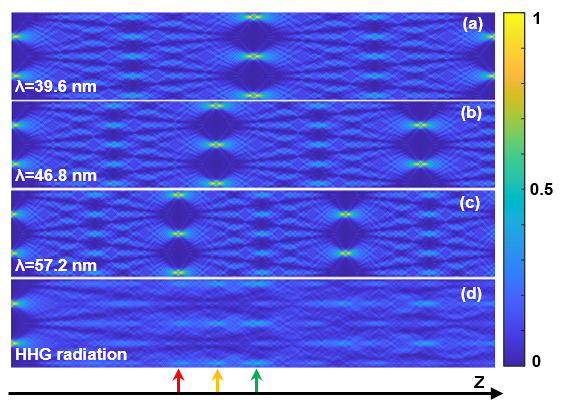}
	\caption{\label{PRL_fig_2} (color online) The distribution field of the Talbot effect along the propagation direction. (a)-(c) are the case of monochromatic plane waves with 39.6 $nm$, 46.8 $nm$, and 57.2 $nm$, respectively. (d) is the case of HHG radiation. The red, orange, and green arrow lines in (d) indicate the self-imaging positions for wavelengths of 57.2 $nm$, 46.8 $nm$, and 39.6 $nm$ in harmonics, respectively.
	}
\end{figure}

HHG beams produce distinct, spatially coherent high-order harmonic spectra, and its diffraction field distribution after passing through the grating can be expressed as:
\begin{eqnarray}
	{u_h}\left( {{{\vec r}_2}} \right) \propto &&\int\limits_{{\lambda _n}} {\int\limits_{{{\vec r}_1}} {\frac{{\exp \left( {i2\pi z/{\lambda _n}} \right)}}{{i{\lambda _n}z}}\exp \left( {i{\phi _n}} \right)} } t\left( {{{\vec r}_1}} \right) \nonumber \\
	&&\times\exp \left[ {i\frac{\pi }{{{\lambda _n}z}}{{\left| {{{\vec r}_2} - {{\vec r}_1}} \right|}^2}} \right]d{{\vec r}_1}\label{eq-5},
\end{eqnarray}
here $\lambda_n$ and $\phi_n$ denote the wavelength and phase of the $n^{th}$-order harmonic beam. Employing Fresnel diffraction theory, we numerically validated the Talbot effect in the HHG beams. For simplicity, a plane wave is assumed. Given the grating with a periodicity of 90 $\mu m$ with an open size of 15 $\mu m$, the wavelengths are 39.6 $nm$, 46.8 $nm$, and 57.2 $nm$ for each harmonic. These parameters are the same as our experiments (see Fig.~\ref{PRL_fig_1}(b)).  Fig.~\ref{PRL_fig_2}(a)-(c) show the distribution of the intensity field along the propagation direction after the monochromatic plane wave passes through the grating. The phenomenon of self-imaging periodically reappears at different orders of Talbot distance is observed. In the case of HHG beams, calculated from Eq.~(\ref{eq-5}), as shown in Fig.~\ref{PRL_fig_2}(d), Diverging from the traditional Talbot effect, the self-imaging phenomenon persists near the $1^{st}$ or $1/2^{th}$ order Talbot distance, aligning with the harmonic's characteristic wavelength. Despite a reduction in contrast due to diffraction superposition from various harmonic orders, the self-imaging effect remains observable.

\begin{figure}[h]
	\includegraphics[width=8.6cm,keepaspectratio]{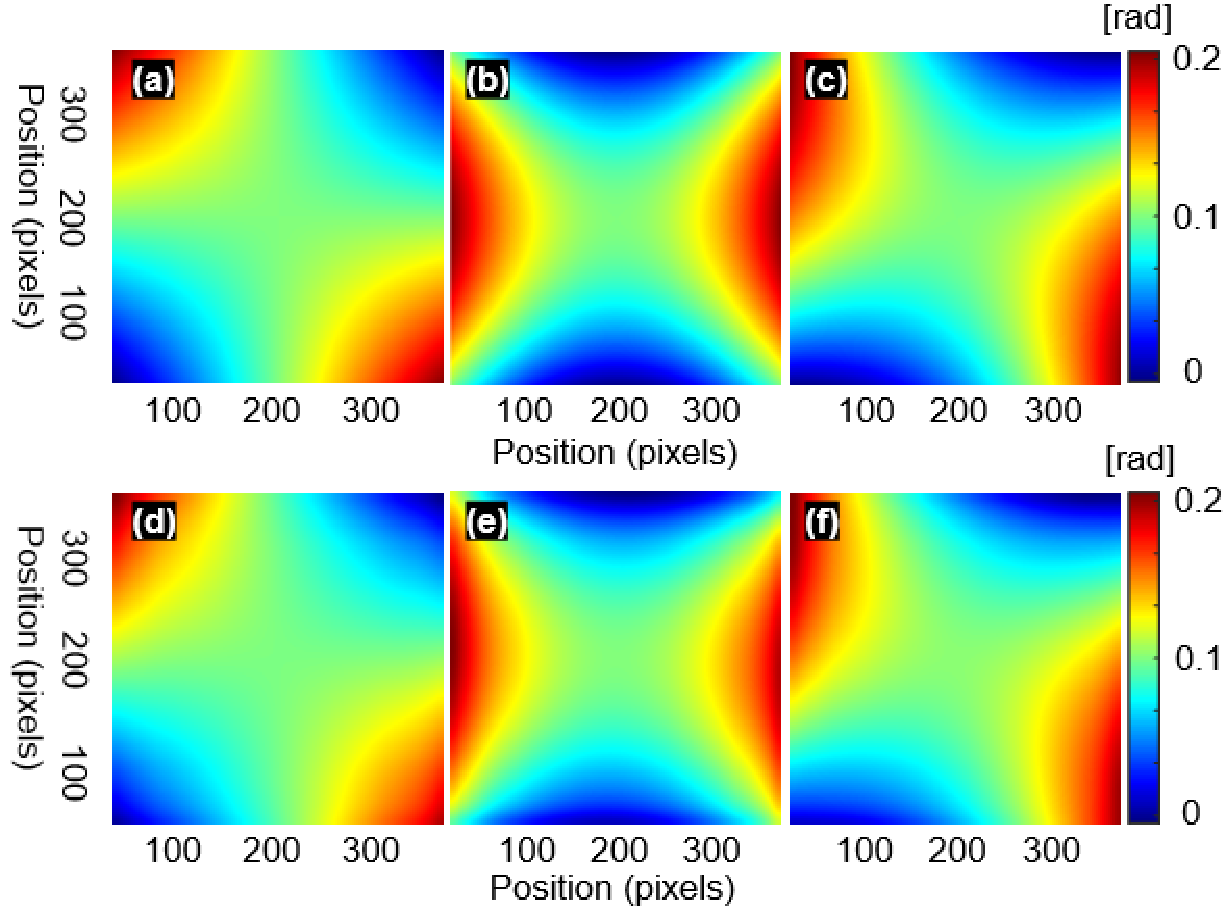}
	\caption{\label{PRL_fig_3} (color online) The simulation results for preserving wavefronts in the Talbot effect of HHG beams. (a)-(c) depict preset wavefronts at wavelengths of 39.6 nm, 48.6 nm, and 57.2 nm, while (d)-(f) show the reconstructed wavefronts obtained from each harmonic's self-imaging.
	}
\end{figure}

 To examine how each harmonic's wavefront propagates in the Talbot effect of HHG beams, we conducted simulations of the wavefront propagation and reconstruction process. Fig.~\ref{PRL_fig_3}(a)-(c) illustrate the preset wavefront of Zernike polynomial components named astigmatism with $z_{2}^{-2}$, $z_{2}^{2}$, and \, $(z_{2}^{-2}+z_{2}^{2})$ \cite{lakshminarayanan2011zernike}, corresponding to the high-order harmonic wavelengths of 39.6 $nm$, 48.6 $nm$, and 57.2 $nm$, respectively. These distances, situated at 204.5 $mm$, 166.7 $mm$, and 141.6 $mm$, correspond to the $1/2^{th}$ Talbot distance of a monochromatic plane wave, capturing the respective self-image of each harmonic. Fig.~\ref{PRL_fig_3}(d)-(f) show the reconstructed wavefronts from individual harmonic Talbot self-images. Their wavefront morphology closely matches the preset wavefronts, demonstrating robust preservation. The reconstruction errors RMS are 0.0496$\lambda$, 0.0384$\lambda$, and 0.0376$\lambda$ for each respective harmonic, underscoring the independence of harmonic wavefront preservation in the Talbot effect. 

A schematic illustration of the experimental configuration is depicted in Fig.~\ref{PRL_fig_1}. The fundamental infrared beam is generated by a commercial 190 fs Yb:KGW femtosecond laser with a center wavelength of 1030 $nm$ ($\textit{Light Conversion, Pharos}$). This laser produces a single pulse energy of 0.92 mJ with a 4 kHz repetition rate, resulting in a 3.68 W average power. To produce a higher flux of about 25 eV harmonics, a driving laser with a central wavelength of 515 $nm$ is generated by bunching the infrared laser to about 5 $mm$ and passing through a 0.5 $mm$ thick BBO crystal. Thus, the driving laser beam with a spot diameter of 5 $mm$, a pulse width of 180 fs and a single pulse energy of 0.45 mJ was obtained. The beam is then passed through two dichroic mirrors to remove the original 1030 $nm$ laser, and the remaining pure 515 $nm$ driving light source is spatially filtered by a half-wave plate and an iris, to expect the highest HHG conversion efficiency and the highest quality EUV spot. A lens with a focal length of $f$ = 300 $mm$ is used to loosely focus the driving condensing light into an argon-filled gas cell featuring a 200 $\mu m$ aperture and a backing pressure of 95 $mbar$. This configuration is characterized as a loosely focused geometry, where the driving laser exhibits a Rayleigh length of 5.7 $mm$ and the beam diameter at a focus of 47 $\mu m$, notably exceeding the estimated 2.5 $mm$ interaction length. Subsequently, two 250 $nm$ thick aluminum films are used to remove the residual driving laser mixed in high-order harmonics. We use photodiode and CMOS to detect pure HHG signals by removing the argon from the gas cell and making sure that our photocurrent drops to zero. The profiles of the spot detected by CMOS at 1130 $mm$ are shown in Fig.~\ref{PRL_fig_1}(b), thus determining that the full angle of HHG beams divergence is about 3 $mrad$. The HHG spectrum after the filtration process is detected by measuring the diffraction from a highly efficient and aberration-free off-plane grating spectrometer \cite{9890574} built by us. By our simulation, the peak power density of the laser at the center of the gas cell is $2.93\times 10^{14}$ W/$cm^{2}$. Considering that the background pressure of argon is 95 $mbar$, the corresponding cutoff photon energy is 32 eV. Due to the cutoff region produced by the 515 $nm$ laser and the high absorption rate of the aluminum film below 17 eV, the obtained harmonics are primarily concentrated in the 39 $nm$-58 $nm$ range, with the dominant harmonics being the $9^{th}$, $11^{th}$, and $13^{th}$ harmonics (corresponding to 57.2 $nm$,46.8 $nm$ and 39.6 $nm$, respectively) shown in Fig.~\ref{PRL_fig_1}, which is highly consistent with our simulations.

The grating employed for the TWS was manufactured on a 25×25 $mm^2$ region using a 10 $\mu m$ thick Tantalum foil. The grating featured a pitch of 90 $\mu m$, and open circle diameters of 15 $\mu m$ were created through femtosecond laser drilling technology, which is a well-known precision manufacturing process with cost-effectiveness and potential for mass production. The calibration and translation of the mask use the piezoelectric linear stage ($\textit{MultiFields Technology}$), the mask underwent scanning along the optical axis, while the diffracted patterns were captured by an sCMOS detector ($\textit{Dhyana XF95, Tucsen Photonics Co., Ltd}$). The detector had dimensions of $2048 \times 2048$ pixels, each with a size of 11 $\mu m$, corresponding to an effective area of 22.5×22.5 $mm^2$. The distance from the focal points to the detector was 1330 $mm$. Following Eq.~(\ref{eq-4}), the $1/2^{th}$ Talbot distances for the $9^{th}$, $11{th}$, and $13^{th}$ harmonics were 161 $mm$, 204 $mm$, and 251 $mm$, respectively, and the self-image of each harmonic is recorded with a single exposure. To determine the displacement vector ($\Delta \xi$, $\Delta \eta$) for the self-image, a reference spot pattern is essential. The reference spots pattern traditionally involved the introduction of a small pinhole in front of the source. However, this approach incurred a significant loss of flux, and the quality of the reference wavefront was notably affected by the morphological characteristics of the pinhole. Additionally, mechanical adjustment errors in the pinhole, such as tilt, pitch, and asymmetry, introduced undesirable effects on the reference wavefront, thereby amplifying the scope and expense of systems necessitating ultra-high vacuum conditions. To address these challenges, we adopted a strategy wherein the spots within the centroid region of the patterns served as the reference. The rationale behind this approach is multifaceted. Firstly, the phase distortion in the central area of the spot is minimal, rendering the resultant spot displacement error negligible. Secondly, this methodology mitigates the limitations associated with pinhole filtering, thereby simplifying system complexity and reducing costs. Ultimately, it enables true reference-free and single-exposure wavefront sensing.

\begin{figure}[h]
	\includegraphics[width=8.6cm,keepaspectratio]{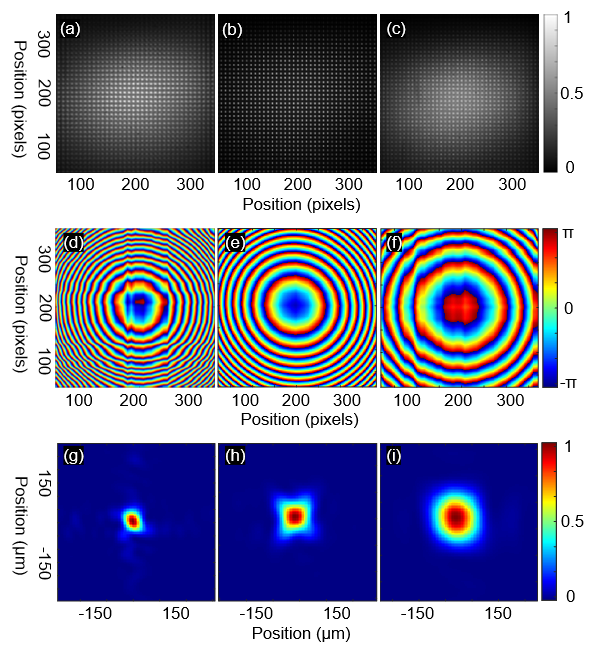}
	\caption{\label{PRL_fig_4} (color online) Experimental results. (a)-(c) represent self-images of high-order harmonic beams at wavelengths 39.6 $nm$, 48.6 $nm$, and 57.2 $nm$.The corresponding (d)-(f) show the wavefront morphology, while (g)-(i) illustrate the focus morphology with Full-width at half-maximum values of 55 $\mu m$, 85 $\mu m$, and 132 $\mu m$, respectively.
	}
\end{figure}

Fig.~\ref{PRL_fig_4}(a)-(c) show the self-images of high-order harmonic beams at wavelengths 39.6 $nm$, 48.6 $nm$, and 57.2 $nm$. These images are captured by the detector with a 100 ms exposure time at the Talbot distances of each harmonic, and the visibilities are 0.2938, 0.8030, and 0.1713, respectively. The shifts of identified spots were calculated using the digital image correlation algorithm with 4 times iterations, achieving a centroid tracking accuracy of $10^{-3}$ pixels \cite{pan2009two,berujon2012two,wang2021nano}. By measuring these varying shifts, the two-dimensional field of the phase distribution can be reconstructed. The shifts of the discerned spots were computed employing the zero-normalized cross-correlation (ZNCC):
\begin{eqnarray}
	\gamma  = \frac{{\sum\limits_{j =  - N}^N {\sum\limits_{j =  - N}^N {\left[ {{I_w}\left( {\xi _i^{'},\eta _j^{'}} \right) - {{\bar I}_w}} \right]\Big[ {{I_r}\Big( {{\xi _i},{\eta _j}} \Big) - {{\bar I}_r}} \Big]} } }}{{\Delta {I_w}\Delta {I_r}}}\label{eq-6},
\end{eqnarray}
where $I_{w}(\xi_{i}^{'}, \eta_{j}^{'})$ represents the intensity in a pixel corresponding to a subset of modulated spots centered at  $(\xi_{0}^{'}, \eta_{0}^{'})$, and  $I_{r}(\xi_{i}, \eta_{j})$ represents the intensity in a pixel of the subset of reference spots centered at  $(\xi_{0}, \eta_{0})$, The terms ${\bar I}_w$ and ${\bar I}_r$ denote the mean values, while $\Delta I_w$ and $\Delta I_r$ represent the standard deviations of the modulated and reference intensities, respectively. The total number of pixels within the traversing window is 2$N$+1. The peak of the cross-correlation coefficient, denoted as $\gamma_{max}$, corresponds to the rigid local displacement $(\Delta \xi, \Delta \eta)$ of the spot in two orthogonal directions, where $\Delta \xi=\xi^{'}-\xi $ and $\Delta \eta=\eta^{'}-\eta$ . The determination of the slope wavefront is attainable through the analysis of individual spot positions $(i, j)$:
\begin{eqnarray}
	\frac{\partial }{{\partial \xi }}\varphi \left( {\xi ,\eta } \right)\left| {_{i,j}} \right. = \frac{{2\pi }}{\lambda }\frac{{{\Delta \xi } \cdot {p_{eff}}}}{{\rm{Z_T}}}\left| {_{i,j}} \right.\label{eq-7},
\end{eqnarray}
and
\begin{eqnarray}
	\frac{\partial }{{\partial \eta }}\varphi \left( {\xi ,\eta } \right)\left| {_{i,j}} \right. = \frac{{2\pi }}{\lambda }\frac{{{\Delta \eta } \cdot {p_{eff}}}}{{\rm{Z_T}}}\left| {_{i,j}} \right.\label{eq-8},
\end{eqnarray}
where the $p_{eff}$ is the effective pixel size in the detector plane, and the overall wavefront can subsequently be derived through an integration process. 

The wavefront is reconstructed by a first-order trapezoidal integration from the slopes, to reduce the accumulated error, the calculations are performed along the forward and reverse directions of the integration path. 
Fig.~\ref{PRL_fig_4}(d)-(f) depict the reconstructed wavefronts for three harmonics, showcased in radians to emphasize fine details beyond curvature. These wavefronts unveil the spherical divergence curvature features as the HHG beam travels through free space. Notably, the main spectral components of the 46.8 $nm$ harmonic exhibit minimal distortion, while the other two harmonics display deviations from the ideal spherical wave. By applying Fresnel back-propagation, the divergence and focal position of each harmonic beam, a comprehensive quantitative examination of the temporal and spatial spectrum and characteristics of the XUV beam's focal point is achievable. At the focal point, as shown in Fig.~\ref{PRL_fig_4}(g)-(i), the spot sizes differ for harmonics with different wavelengths, with approximate half-widths of 55 $\mu m$, 85 $\mu m$, and 132 $\mu m$, respectively. This demonstrates another crucial aspect of the attosecond XUV pulse focusing characteristics, namely, wavelength-dependent imaging dispersion in high-order harmonic beams. Previous studies have explored how aberrations transfer from the fundamental wave to HHG beams, However, 
prior reported wavefront measurements with restricted spatial resolution, relying on spectral averaging or spectral-resolved wavefront measurements involving multiple exposures and extensive iterative calculations. Our proposed spectral resolved TWS technique offers a solution to these limitations, potentially leading to new discoveries in the mechanism of physical interactions.

In conclusion, we have discovered a unique phenomenon in the HHG Talbot effect, highlighting the preservation of high-order harmonic wavefronts. This novel effect provides a groundbreaking method for precisely assessing the spectral resolution of broadband high-order harmonic wavefronts. With its notable attributes of high sensitivity and precision, friendly to large NA, and compatibility with single exposures, this discovery holds great potential in multi-color wavefront metrology. Particularly, it breaks the constrain of quasi-monochromatic waves or wavelength-averaging in the conventional TWS, extends the spectrum-resolved wavefront diagnostic capability, and presents exciting possibilities for detecting shot-by-shot variations in the HHG beams. Looking ahead, this effect opens avenues for exploring complex aberration transfer mechanisms in the HHG process, potential spatiotemporal coupling in attosecond pulses, wavefront shaping of XUV beams, and diverse applications across multiple disciplines.

This work was funded by the Shenzhen Science and Technology Program (JCYJ20220530140805013, KQTD20170810110313773), the National Key Research and Development Program of China (2021YFB3602600), the Chinese Academy of Sciences (GJJSTD20200009, 2018-131-S), and the National Natural Science Foundation of China (12074167, 62121003, 10010108B1339-2451). We wish to thank Henry N Chapman and Sa$\rm \check{s}$a Bajt at DESY for their guidance on wavefront sensing, and Lifeng Wang at IASF for his crucial discussions.

\nocite{*}

\bibliography{apssamp}

\end{document}